\documentclass[aps, prd, preprint, amsfonts, floatfix]{revtex4}
\usepackage{hyperref}

\usepackage{graphicx}
\usepackage{amsmath,amsfonts}
\usepackage{epsfig,color}

\begin{document}
\newcommand{\be}{\begin{eqnarray}}
\newcommand{\ee}{\end{eqnarray}}
\newcommand\del{\partial}
\newcommand\nn{\nonumber}
\newcommand{\Tr}{{\rm Tr}}
\newcommand{\Str}{{\rm Trg}}
\newcommand{\mat}{\left ( \begin{array}{cc}}
\newcommand{\emat}{\end{array} \right )}
\newcommand{\vect}{\left ( \begin{array}{c}}
\newcommand{\evect}{\end{array} \right )}
\newcommand{\tr}{{\rm Tr}}
\newcommand{\hm}{\hat m}
\newcommand{\ha}{\hat a}
\newcommand{\hz}{\hat z}
\newcommand{\hze}{\hat \zeta}
\newcommand{\hx}{\hat x}
\newcommand{\hy}{\hat y}
\newcommand{\tm}{\tilde{m}}
\newcommand{\ta}{\tilde{a}}
\newcommand{\U}{\rm U}
\newcommand{\diag}{{\rm diag}}
\newcommand{\tz}{\tilde{z}}
\newcommand{\tx}{\tilde{x}}
\definecolor{red}{rgb}{1.00, 0.00, 0.00}
\newcommand{\rd}{\color{red}}
\definecolor{blue}{rgb}{0.00, 0.00, 1.00}
\definecolor{green}{rgb}{0.10, 1.00, .10}
\newcommand{\blu}{\color{blue}}
\newcommand{\green}{\color{green}}



\title{Subsets and the canonical partition functions}

\author{Jacques Bloch}
\author{Falk Bruckmann}
\affiliation{Institute for Theoretical Physics, University of Regensburg, 
93040 Regensburg, Germany}

\author{Mario Kieburg}
\affiliation{Department of Physics and Astronomy, SUNY${{}}$, Stony Brook,
 New York 11794, USA}

\author{K. Splittorff}
\affiliation{Discovery Center, The Niels Bohr Institute, University of Copenhagen, 
Blegdamsvej 17, DK-2100, Copenhagen {\O}, Denmark} 
\author{J.J.M. Verbaarschot}
\affiliation{Department of Physics and Astronomy, SUNY${  }$, Stony Brook,
 New York 11794, USA}

\date   {\today}
\begin  {abstract}
We explain the physical nature of the subset solution to the sign problem 
in chiral random matrix theory: The subset sum is shown to project out the 
canonical determinant with zero quark charge from a given configuration. 
As the grand canonical chiral random matrix partition function is independent 
of the chemical potential, the zero quark charge sector provides the full result.
\end{abstract}
\maketitle

\section{Introduction}

Chiral random matrix theory \cite{chRMT} has given us several deep insights 
into the QCD sign problem which prohibits direct application of 
lattice QCD methods at nonzero quark chemical potential \cite{signRev}. 
The random matrix framework has allowed us to 
understand the failure of the quenched approximation \cite{Misha}, to 
formulate the OSV relation \cite{OSV} which replaces the Banks-Casher 
relation \cite{BC} and to derive the first analytical 
result for the average 
phase factor of the fermion determinant \cite{SV-phase}.   
 
These lessons from chiral random matrix theory apply directly to 
QCD at nonzero chemical potential since the two are equivalent 
in the microscopic limit or the 
$\epsilon$-regime of chiral perturbation theory.   
For these reasons it is most interesting that the sign problem in a chiral 
random matrix theory can be solved by means of the subset method 
\cite{Bloch1,Bloch2}. In particular this subset method works even 
in the region of $\mu>m_\pi/2$ where the sign problem is severe.

The aim of the present paper is to provide the 
physical explanation of why the subset method introduced in 
\cite{Bloch1,Bloch2} solves the sign problem in chiral random matrix 
theory: 
As we will show in detail below the subset construction projects 
out the canonical determinant with zero quark charge from the fermion 
determinant. Since the chiral random matrix partition function is 
independent of the chemical potential the zero charge part makes up 
the full result.

In this paper we start from a random matrix theory for QCD that is 
$\mu$-independent even for finite size of the random matrix. This choice 
has a direct physical motivation: First, in the microscopic domain 
 (where the size, $n$, of the random matrix goes 
to infinity while the quark mass times $n$ and the square of the chemical 
potential times $n$ are held fixed) the random matrix partition function  
is identical to the partition function of 
chiral perturbation theory in the $\epsilon$-regime \cite{Split-V-factor}. 
Second, being a theory
of pions, which are bound states of quarks and anti-quarks, chiral 
perturbation theory naturally does not couple to the quark chemical 
potential. As chiral perturbation theory is the effective theory for QCD 
at low temperatures and $\mu <m_N/3$ the $\mu$-indenpendence of the partition 
function is natural in this regime.

In general, random matrix partition functions for QCD only need to be
$\mu$-independent in the microscopic domain. In the present context 
it is convenient to work with a random matrix theory where this
$\mu$-independence is manifest even at finite $n$.

\section{Subset and canonical determinants in chRMT}

Our starting point is a variation of the  chiral Random Matrix Theory 
(chRMT) at non-zero chemical potential $\mu$ introduced in \cite{O} 
(see \cite{Akemann:2007rf} for a review). It is defined by
\be
Z(m,\mu) =\int d\Phi_1 d\Phi_2  
\det\left(\begin{array}{cc} m & \hspace{-5mm} e^\mu \Phi_1 - e^{-\mu} \Phi_2^\dagger \\ 
-e^{-\mu} \Phi_1^\dagger+e^\mu \Phi_2 & \hspace{-5mm} m \end{array}\right) 
e^{-n\Tr(\Phi_1\Phi_1^\dagger+\Phi_2\Phi_2^\dagger)}, 
\label{Z_mu-indp}
\ee
where $\Phi_1$ and $\Phi_2$ are complex $n\times n$ matrices. 
We have chosen 
to work with this form of the partition function because it is independent of 
$\mu$ even for finite $n$ (this was also the case for the partition
function used in \cite{O}), and because the chemical potential appears in the
form $\exp(\pm \mu)$ which allows us to project out the canonical partition
function in the same way as in lattice QCD. The $\mu$-independence of the
partition function follows immediately by using that the Gaussian integral
is only nonzero for terms that have an equal number of factors $\Phi_i$ and
$\Phi^\dagger_i$ for $ i= 1, \, 2$.
The relation to the form used in \cite{Bloch1,Bloch2} is given in the appendix. 
 
In the subset method of \cite{Bloch1,Bloch2}, one first performs   a sum over  a subset of roots of unity contained
in the integral over  the matrices $\Phi_1$ and $\Phi_2$. The critical observation in  \cite{Bloch1,Bloch2} 
that the determinants
\be
d(\mu,\theta_k) \equiv
\det\left(\begin{array}{cc} m & 
\hspace{-5mm} 
 e^{\mu+i\theta_k} \Phi_1 -e^{-\mu-i\theta_k} \Phi_2^\dagger \\
  -e^{-\mu-i\theta_k}\Phi_1^\dagger + e^{\mu+i\theta_k}\Phi_2
 & \hspace{-5mm} m \end{array}\right),
\label{det_sub} 
\ee
where $\theta_k={2 k\pi}/{N_s}$ with $N_s\geq 2n+1$ \footnote{Due to chiral symmetry one can also instead use $\theta_k={k\pi}/{N_s}$ with $N_s\geq n+1$ \cite{Bloch1,Bloch2}. For numerical implementations this gains a factor of 2.} sum up to a positive real number 
\be
\frac{1}{N_s}\sum_{k=0}^{N_s-1} d(\mu,\theta_k) \in {\bf R}_+.
\label{measure}
\ee
This number can then in turn be used to generate a Monte Carlo ensemble of 
configurations and subsequently the unquenched expectation values. 
Note the invariance of the Gaussian measure under these 
phase rotations (the arguments below apply to any measure with the same invariance properties). Next we show that the measure  Eq. (\ref{measure})
is a canonical determinant with zero baryon number which is manifestly
positive. 

For a given configuration $(\Phi_1,\Phi_2)$ we  decompose the fermion determinant 
\be
D(\mu) \equiv 
\det\left(\begin{array}{cc} m & \hspace{-5mm} e^\mu \Phi_1 - e^{-\mu} \Phi_2^\dagger \\ 
-e^{-\mu} \Phi_1^\dagger+e^\mu \Phi_2 & \hspace{-5mm} m \end{array}\right) 
\ee
into canonical determinants
\be
D(\mu) = \sum_{q=-2n}^{2n} e^{\mu q}D_q,
\ee
where 
\be
D_q \equiv \frac{1}{2\pi}\int_{-\pi}^\pi d\theta \ e^{-iq\theta} D(i\theta).
\ee
(See \cite{BDGLL,DG,Alexandru:2010yb} for applications of canonical determinants to lattice QCD.)
Likewise we decompose the partition function, $Z$, into canonical 
partition functions
\be
Z(\mu) = \sum_{q=-2n}^{2n} e^{\mu q}Z_q,
\ee
where ($\langle\ldots\rangle$ is the expectation value with respect to the 
Gaussian weight for $\Phi_{1,2}$)
\be
Z_q =\langle D_q \rangle.
\ee
As $Z$ is independent of $\mu$ we necessarily  have $Z_q=0$ for $q\neq0$. 
For odd $q$ the canonical determinants vanish as well, $D_{q=2l+1} = 0$.
This follows trivially from $D(i(\mu+\pi))=D(i\mu)$ and 
$\exp(-iq(\mu+\pi))=\exp(-iq\mu)(-1)^q$.
For even index, however, the canonical determinants are nonzero 
for a typical configuration $(\Phi_1,\Phi_2),$ and only after 
averaging will one find $Z_{q=2l}= 0$ for $l\neq0$. 

To make the connection to the subset construction of \cite{Bloch1,Bloch2} we 
first rewrite the canonical partition functions
\be
D_q 
& = & \frac{1}{2\pi}\int_{-\pi}^\pi d\theta \ e^{-iq(-i\mu+\theta)} D(i(-i\mu+\theta)) \nn \\
& = & \frac{1}{2\pi}e^{-q\mu}\int_{-\pi}^\pi d\theta \ e^{-iq\theta} D(\mu+i\theta),
\label{Dq_v2}
\ee
where in the first line we shifted the contour into the complex plane.
The determinant inside the integrand is now
\be
D(\mu+i\theta) & = &  
\det\left(\begin{array}{cc} m & \hspace{-5mm} e^{\mu+i\theta} \Phi_1 
- e^{-\mu-i\theta} \Phi_2^\dagger \\ 
-e^{-\mu-i\theta} \Phi_1^\dagger+e^{\mu+i\theta} \Phi_2 & \hspace{-5mm} m \end{array}\right) .
\label{Dmu_itheta}
\ee
\vspace{3mm}

To establish the relation between this subset construction and the 
canonical determinants introduced above first note that we can 
replace the subset-sum over $\theta_k$ in Eq.~(\ref{measure}) by an integral
\be
\frac{1}{N_s}\sum_{k=0}^{N_s-1} d(\mu,\theta_k) 
= \frac{1}{2\pi}\int_{-\pi}^\pi d\theta \ d(\mu,\theta).
\ee
This follows from the observation that the integrand is a polynomial in 
$e^{\pm i\theta}$ of maximum order $2n$ and that all integrals follow from 
the orthogonality relations
\be
\frac 1{2\pi}\int_{-\pi}^\pi d\theta \ e^{i(j-l) \theta} = \delta_{jl}.
\ee
The same orthogonality relation holds for the sum 
\be
\frac{1}{N_s}\sum_{q=0}^{N_s-1} e^{i\theta_q(j-l)} =\delta_{jl}
\ee
provided that $|j|,|l| \le N_s $. Therefore the sum over $k$ gives the exact 
value of the integral if $N_s \ge 2n+1$.

By comparison of Eq.~(\ref{det_sub}) with Eq.~(\ref{Dmu_itheta}) we then 
see that the subset sum is equivalent to the projection onto the $q=0$ 
canonical determinant, that is
\be
\frac{1}{N_s}\sum_{k=0}^{N_s-1} d(\mu,\theta_k) = D_0.
\ee
This is the physical explanation of what the subset is. 

Since $Z_{q=0}=Z$ (the $q=0$ part makes up the entire partition function 
because it is independent of $\mu$) the subset method 
gives the full result.   
Moreover, as the subset sum for a given configuration is equivalent to the 
canonical determinant with $q=0$,  it is clear that the subset sum is necessarily 
real and positive: as can be seen explicitly from Eq.\,(\ref{Dq_v2}), we have 
that $D_{q=0}$ is independent of $\mu$, and for $\mu=0$ all determinants 
in the subset sum are real and positive. This is the physical explanation 
of why the subset method works, see also \cite{Bloch1,Bloch2}. 

For the variant of the chiral random matrix partition function used in 
\cite{Bloch1,Bloch2} the interpretation of the subset is analogous, see 
\ref{app:reformulate}.  The original argument for why the subset method 
works given in \cite{Bloch1,Bloch2} is also related to the argument given 
above.

In general the QCD partition function will of course depend on the chemical 
potential and hence in QCD one will need to evaluate all $D_q$. 
For the evaluation of $D_q$ it is also possible to turn the 
integral into a sum. In this case the maximum order of the polynomial in 
$e^{\pm i\theta}$ is $2n+|q|$, and therefore the subset sum evaluates the integral exactly for $N_s \ge 2n+|q|+1$.  
Note, however, that the $D_q$ with $q\neq0$ are not real and positive so 
we do not have a  weight to perform Monte Carlo Simulations (even if we can
do these integrals 
exactly). This is in exact analogy with the observations of 
\cite{Alexandru:2010yb} in lattice QCD.

The argument given above also applies if $\Phi$ is unitary rather than 
complex. More generally, for unitary lattice gauge theories where the chemical  
potential is introduced into the temporal links by \cite{HK}
\be
U_t  &\to& e^\mu U_t, \nn \\
U_t^\dagger &\to& e^{-\mu} U_t^\dagger .
\ee
the partition function is $\mu$-independent and equal to the charge
zero canonical partition function. The subset method then applies in
exactly the same way as in the random matrix model discussed above.

\section{Conclusions}
\label{sec:conc}

The subset solution to the sign problem in chiral random matrix theory 
has been shown to be equivalent to the projection, configuration by 
configuration, onto the zero quark number canonical determinants. Since 
the chiral random matrix partition function is independent of the 
chemical potential, the canonical partition function makes up the full
grand canonical partition function. This gives the physical reason how 
the subset construction works. The same argument applies to unitary 
lattice gauge theories at nonzero chemical potential.

The vanishing value of the canonical partition functions in chiral random 
matrix theory for nonzero quark number is the result of detailed 
cancellations: the canonical determinants with nonzero quark number 
take complex values and only the average value is zero. 
The projection onto the canonical 
determinants with nonzero quark number can also be obtained from a subset 
sum, however, it remains a challenge to devise a numerical method to control 
the cancellations in the average. Such a method would potentially have direct 
application to full QCD where partition functions with $q\neq 0$ are 
nonvanishing. It may also be able to cast further light on the special nature 
of the noise \cite{Kaplan1,Kaplan2} related to the sign problem.

Despite the $\mu$-independence of the chiral random matrix partition function
the random matrix theory gives a plethora of nontrivial results for the 
spectral correlation functions of the Dirac operator and for the fluctuations
of the fermion determinant. The reason for this is that the generating 
functionals for such partially quenched observables have a highly nontrivial 
dependence on the chemical potential. 
It would be most interesting if one would be able to extend the subset 
method to these partially quenched observables.  

\noindent
{\bf Acknowledgments:}
We would like to thank Christof Gattringer for discussions and exchange 
of unpublished results. 
This work was supported by U.S. DOE Grant No. DE-FG-88ER40388 (JV), 
the {\sl Sapere Aude} program of The Danish Council for 
Independent Research (KS), the DFG collaborative research
center SFB/TR–55 (JB), the DFG (BR 2872/4-2) (FB) and the 
Alexander-von-Humboldt Foundation (MK).


\renewcommand{\thesection}{Appendix \Alph{section}}
\setcounter{section}{0}

\section{Equivalence of the chRMT formulations}
\label{app:reformulate}

The form of the chRMT used in \cite{Bloch1,Bloch2} was
\be
Z_B(m,\mu_B) = \int d\Phi_1 d\Phi_2 \det\left(\begin{array}{cc} m & i\Phi_1+\mu_B\Phi_2 \\ i\Phi_1^\dagger+\mu_B\Phi_2^\dagger & m \end{array}\right) e^{-n\Tr(\Phi_1\Phi_1^\dagger+\Phi_2\Phi_2^\dagger)}.
\label{ZmuB}
\ee
This partition function depends on $\mu_B$ for finite $n$ \cite{O}. Because 
of the $\mu_B$-dependence of the partition function $Z_B(m,\mu_B)$, the subset 
sum is not equal to the canonical partition function
for $q_B= 0$ and the corresponding canonical partition functions for 
$q_B\ne 0$ are nonvanishing. The $\mu_B$-dependence is, however, of a form 
where the partition function at non-zero $\mu_B$ is trivially related 
to the one at $\mu_B=0$ \cite{O}
\be
\label{muBdep}
Z_B(m,\mu_B) = (1-\mu_B)^n Z_B(\frac{m}{\sqrt{1-\mu_B^2}},0).
\ee
In \cite{Bloch2} it was shown that the subset sum for each configuration 
realizes this relation. When $\mu_B<1$ both the prefactor $(1-\mu_B)^n$ 
and the rescaled quark mass are real and positive thus, as originally 
argued in \cite{Bloch1,Bloch2}, the subset sum for the right hand side 
is always real and positive. The relation, Eq.~(\ref{muBdep}), is the 
analogue of the $\mu$-independence of the chRMT used in this paper, and 
the fact that subsets realizes this relation configuration by configuration 
is the analogue of the projection onto the canonical determinant with 
zero quark charge. 

The form of the chRMT adopted in Eq.\,(\ref{Z_mu-indp}) is related to 
the form, Eq.~(\ref{ZmuB}), used in \cite{Bloch1,Bloch2} by a $\mu$ dependent rescaling 
of the mass and a trivial overall factor. If we start from  
\be
Z(m,\tilde\mu) = \frac{1}{(1-\tilde\mu^2)^n} \int d\Phi_1 d\Phi_2 \det\left(\begin{array}{cc} m\sqrt{1-\tilde\mu^2} & i\Phi_1+\tilde\mu\Phi_2 \\ i\Phi_1^\dagger+\tilde\mu\Phi_2^\dagger & m\sqrt{1-\tilde\mu^2} \end{array}\right) e^{-n\Tr(\Phi_1\Phi_1^\dagger+\Phi_2\Phi_2^\dagger)},
\label{Ztildemu}
\ee
then it is clear from Eq.~(\ref{muBdep}) that $Z(m,\tilde\mu)$ is 
independent of $\tilde\mu$. Moreover, with $\tilde\mu$ given by 
\be
\tanh(\mu) = \tilde\mu,
\ee
then this partition function is identical to the one of Eq.\,(\ref{Z_mu-indp}).
In order to see this first note that $\cosh(\mu)=1/\sqrt{1-\tilde\mu^2}$ and 
$\sinh(\mu)=\mu/\sqrt{1-\tilde\mu^2}$ and then use this to express the 
determinant in terms of $\mu$ 
\be
&& \det\left(\begin{array}{cc} m/\cosh(\mu) & 1/\cosh(\mu)(i\cosh(\mu)\Phi_1+\sinh(\mu)\Phi_2) \\ 1/\cosh(\mu)(i\cosh(\mu)\Phi_1^\dagger+\sinh(\mu)\Phi_2^\dagger) & m/\cosh(\mu) \end{array}\right) \nn \\
&=& 1/\cosh^{2n}(\mu) \det\left(\begin{array}{cc} m & i\cosh(\mu)\Phi_1+\sinh(\mu)\Phi_2 \\ i\cosh(\mu)\Phi_1^\dagger+\sinh(\mu)\Phi_2^\dagger & m \end{array}\right). 
\ee
The factor $1/\cosh^{2n}(\mu)$ cancels against the prefactor $1/(1-\tilde\mu^2)^n$ in the partition function of Eq.~(\ref{Ztildemu}). After choosing 
\be
\Phi_1' =\frac i2 (\Phi_1-i\Phi_2), \qquad \Phi_2' =\frac i2( \Phi_1^\dagger -i\Phi_2^\dagger) 
\ee
as new integration variables, we recover
the form given in Eq.~(\ref{Z_mu-indp}).

The subsets defined in \cite{Bloch1,Bloch2} consist of rotated matrices
\begin{equation}
\Phi_1 \to \cos\theta_k\Phi_1+\sin\theta_k\Phi_2, \quad \quad
\Phi_2 \to -\sin\theta_k\Phi_1+\cos\theta_k\Phi_2
\end{equation}
which translates into
\begin{equation}
\Phi'_{1,2} \to e^{i\theta_k}\Phi'_{1,2},
\end{equation}
as in Eq.~(\ref{det_sub}).



\end{document}